\documentclass[12pt,preprint]{aastex}
\usepackage{multirow,amsthm,amsmath,amssymb,mathrsfs,ulem}
\usepackage{graphicx,subfigure,verbatim,color,soul,lineno}

\graphicspath{{./}}

\begin{document}

\title{The AMS-02 cosmic ray deuteron flux is consistent with a secondary origin}

\author{Qiang Yuan$^{1,2}$, Yi-Zhong Fan$^{1,2}$}

\affil{$^1$Key laboratory of Dark Matter and Space Astronomy, Purple
Mountain Observatory, Chinese Academy of Sciences, Nanjing 210023, China;
yuanq@pmo.ac.cn, yzfan@pmo.ac.cn\\
$^2$School of Astronomy and Space Science, University of Science and
Technology of China, Hefei 230026, Anhui, China}

\begin{abstract}
The recent measurements of cosmic ray deuteron fluxes by AMS-02 show that 
the rigidity dependence of deuterons is similar with that of protons but 
flatter than $^3$He, which has been attributed to the existence of primary 
deuterons with abundance much higher than that from the Big Bang 
nucleosynthesis. The requirement of highly deuteron-abundant sources 
imposes a serious challenge on the modern astrophysics since there is 
no known process to produce a large amount of deuterons without violating 
other constraints \citep{1976Natur.263..198E}. In this work we demonstrate 
that the fragmentation of heavy nuclei up to nickel plays a crucial role 
in shaping/enhancing the spectrum/flux of the cosmic ray deuterons. 
Based on the latest cosmic ray data, the predicted secondary fluxes 
of deuterons and $^3$He are found to be reasonably consistent with the 
AMS-02 measurements and a primary deuteron component is not needed. 
The observed differences between the spectra of D and $^3$He, 
as well as those between the D/$^4$He (D/p) and $^3$He/$^4$He ($^3$He/p) 
flux ratios, measured in the rigidity space, is probably due to the 
kinetic-energy-to-rigidity conversion and the solar modulation, 
given different charge-to-mass ratios of D and $^3$He. 
More precise measurements of the fragmentation cross sections of various 
nuclei to produce deuterons, tritons, and $^3$He in a wide energy range 
will be very helpful in further testing the secondary origin of cosmic 
ray deuterons.
\end{abstract}

\keywords{cosmic rays}

\section{Introduction}
In the current standard cosmological model, the lightest elements, including 
H, D, $^{3}$He, $^{4}$He, and $^{7}$Li, were produced in the first few 
minutes of the Big Bang \citep{1998RvMP...70..303S}. Observations of D/H 
can limit or measure the intrinsic primordial abundance because D is thought 
to be destroyed by stars and thus D/H should monotonically decrease after 
the Big Bang nucleosynthesis. Various measurements suggest that 
D/H$\approx 2.5\times 10^{-5}$ in the Universe \citep{2003ApJ...597...48P}. 
However, among the cosmic rays (CRs), deuterons are much more abundant 
(i.e., approximately $2\%-3\%$ of the proton abundance) and such particles 
are widely attributed to the spallation of the heavy CR nuclei 
\citep{2002ApJ...564..244W,2016ApJ...818...68A}. Predictions from the CR 
modeling were found to be roughly consistent with the state-of-the-art 
measurements \citep{2012A&A...539A..88C,2023PhRvD.107l3008G}. However, 
relatively low energy coverage and large uncertainties of the measurements 
hinder a decisive test of the origin of CR deuterons.

Very recently, the AMS-02 collaboration published their accurate measurements 
of deuteron fluxes, which are based on 21 million D nuclei in the rigidity 
range from 1.9 to 21 GV collected from May 2011 to April 2021 
\citep{PhysRevLett.132.261001}. The spectral behaviors of the D and $^{3}$He 
CRs are found to be very different, which has been interpreted as the 
presence of a primary-like deuteron component consisting of $(9.4\pm 0.5)\%$ 
of the $^{4}$He flux \citep{PhysRevLett.132.261001}. If significant spallation 
takes place in CR acceleration sources, the produced deuterons will have a 
hard spectrum similar with or even harder than (in case of further acceleration
of secondary particles) that of primary CRs \citep{2009PhRvL.103e1104B}. 
This possibility, however, will be strongly challenged by the consistency 
of the $^{3}$He data with the secondary origin model. Likely, unless the 
acceleration of deuterons are about $10^2-10^3$ times more efficient than 
that of protons, a non-primordial origin of a significant amount of deuterons 
in CR sources is needed. 

Motivated by the considerably enhanced D/H found in Orion Nebula and some 
low temperature molecular clouds, some non-primordial origin models for 
deuterium have been proposed in the literature 
\citep{1973Natur.241..384H,1999ApJ...511..502M}. The challenge is that the 
deuterium is usually destroyed in the astrophysical processes. The 
astrophysical production of such particles in principle has not been 
definitely ruled out, but the conditions are usually too severe to be 
satisfied \citep{1976Natur.263..198E}. It was argued that in flaring 
dwarf stars, significant non-primordial deuterium could arise as a 
secondary product of neutrons in stellar flares which then capture on 
protons via $n+p\rightarrow {\rm D}+\gamma$ \citep{1999ApJ...511..502M}. 
For reasonable flare spectra, the dedicated calculation found that 
$n/{\rm D}\leq 10$ and $(n+{\rm D})/^{6}{\rm Li}\leq 400$, which can not 
serve as an important Galactic source of deuterium \citep{2003ApJ...597...48P}. 

In summary, so far, there is no viable channel that is able to produce a 
large amount of non-primordial deuterium. If there is indeed a significant 
primary-like CR deuteron component in the AMS-02 data, as speculated in 
\citet{PhysRevLett.132.261001}, ``new" astrophysics or even ``new" physics 
will be called for. Therefore, it is crucial to clarifying whether there 
is indeed a  deuteron excess on top of the secondary production in the 
current data. Motivated by such a fact, here we carry out a dedicated 
calculation on the CR deuteron production and report a spectrum consistent 
with the AMS-02 data.  
  
\section{Deuteron flux from secondary production} 

We use the numerical tool GALPROP \citep{1998ApJ...509..212S} v56 to calculate 
the production and propagation of secondary particles. The framework of 
diffusion propagation with stochastic reacceleration in the interstellar 
medium (ISM), which is found to well reproduce the boron-to-carbon ratio 
(B/C) of CRs, is assumed \citep{2017PhRvD..95h3007Y,2020JCAP...11..027Y}. 
The diffusion coefficient was usually parameterized as a power-law of 
particle rigidity, and the power-law index is related with the property 
of the ISM turbulence. The recent measurements of CR secondary-to-primary 
ratios by AMS-02 and in particular DAMPE show clear hardenings at $\sim100$ 
GeV/n \citep{2021PhR...894....1A,2022SciBu..67.2162D}. We therefore 
introduce a high-energy break of the rigidity dependence of the diffusion 
coefficient. The reacceleration effect can be described by a diffusion in 
the momentum space, and the Alfven velocity ($v_A$) of the magnetohydrodynamic 
waves is used to characterize the reacceleration effect 
\citep{1994ApJ...431..705S}. As for the injection spectra of primary CRs, 
we use a non-parametric, spline interpolation method to describe them 
\citep{2018ApJ...863..119Z}. The spectral parameters are assumed to be 
the same for all nuclei but protons and helium. See Appendix A for more 
details about the determination of the propagation parameters and the 
injection spectra.

\begin{figure*}[!htb]
\centering
\includegraphics[width=0.48\textwidth]{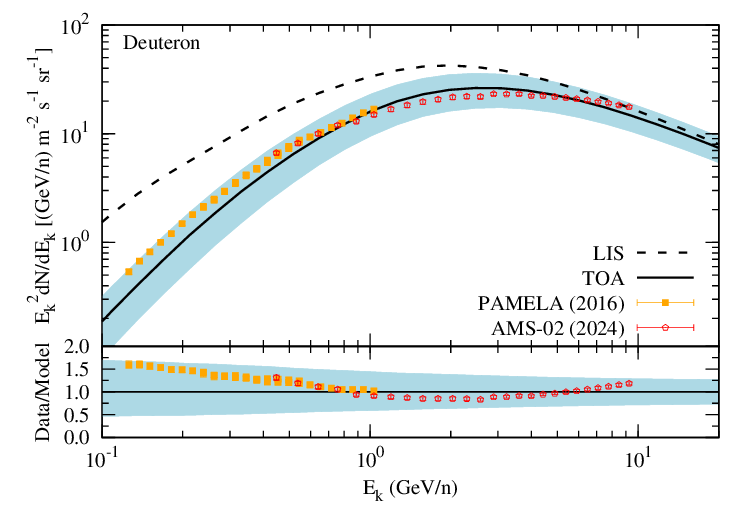}
\includegraphics[width=0.48\textwidth]{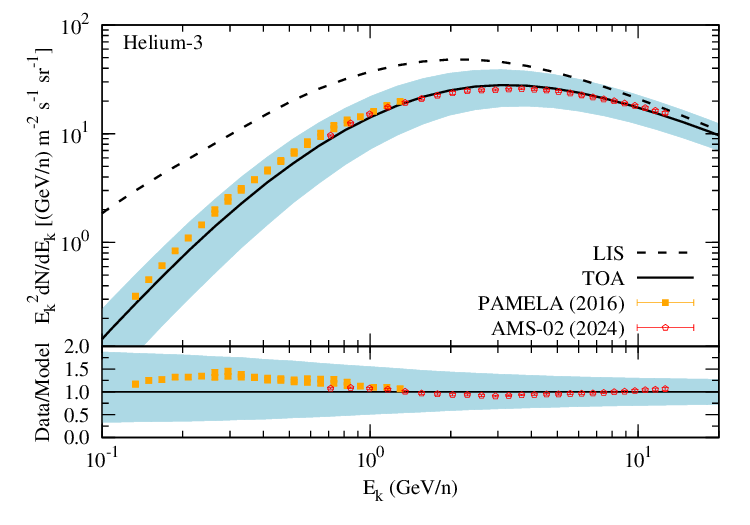}
\caption{Expected fluxes of D (left) and $^3$He (right) from fragmentation of 
all nuclei, compared with the measurements by PAMELA \citep{2016ApJ...818...68A} 
and AMS-02 \citep{PhysRevLett.132.261001}. The dashed line is the local interstellar 
(LIS) spectrum, the solid line corresponds to the modulated spectrum on top of 
the atmosphere (TOA) for $\phi=600$ MV, and the shaded bands show the uncertainty 
ranges of the model predictions. The bottom sub-panels show the data-to-model 
ratios together with the uncertainty bands of the predictions.}
\label{fig:dhe3_DRBR}
\end{figure*}

The fragmentation cross section is crucial to correctly calculating the 
yield fluxes of secondary particles 
\citep{2018PhRvC..98c4611G,2024PhRvC.109f4914G}. We note that in the
v56 version of GALPROP, most of the cross sections to produce D, T, and
$^3$He from fragmentations of nuclei heavier than oxygen are lack. In the
latest v57 release of the code, more cross sections have been added
\citep{2022ApJS..262...30P}. We implement the new parameterizations of 
production cross sections of D, T, and $^3$He up to nickel nuclei 
\citep{2012A&A...539A..88C} into the GALPROP v56 code, with which our 
propagation parameters were obtained. A similar approach was used 
in \citet{2023PhRvD.107l3008G} to provide the GALPROP calculation shown 
in the AMS-02 publication\footnote{However, it seems that the D pruduction
cross section in the implementation of \citet{2023PhRvD.107l3008G} 
is lower than that of \citet{2012A&A...539A..88C}.}. 
Fragmentations of heavy nuclei were also 
found to be important for the lithium production, but affect little the 
boron production \citep{2022A&A...668A...7M}. Furthermore, the existing 
cross sections for the major channel of helium fragmentation (including 
$^3$He$\to$D) are updated or added. There is also a channel to produce 
deuterons from proton-proton fusion, $p+p\to D+\pi^+$. The peak cross 
section of this fusion process is about 3 mb for proton kinetic energy 
of $\sim0.6$ GeV \citep{1972A&AS....7..417M}. Due to the high fluxes of 
protons compared with other nuclei, this process also contributes to the 
production of deuterons even if the cross section is small. 

The calculated fluxes of D and $^3$He from fragmentation of nuclei up to 
$Z=28$ (nickel), together with the measurements by PAMELA 
\citep{2016ApJ...818...68A} and AMS-02 \citep{PhysRevLett.132.261001}, 
are shown in Fig.~\ref{fig:dhe3_DRBR}. To compare with the measurements 
at low energies, we employ the force-field approximation of the solar 
modulation effect \citep{1968ApJ...154.1011G}. The modulation potential 
derived from the fitting in Appendix A is about 600 MV. However, we take 
a range of the modulation potential of $600\pm150$ MV to consider the 
difference among the data-taken time of PAMELA (2006-2007) and AMS-02 
(2011-2021) for the deuteron measurements and those used for the propagation
and injection parameter fitting (2011-2018). We further include the 
uncertainties from the propagation parameters (estimated to be $\sim15\%$;
see Appendix A) and the cross sections to produce D and $^3$He (estimated 
to be about $10\%$ for D and $\sim 10-20\%$ for $^3$He; 
\citet{2023PhRvD.107l3008G}). All these uncertainties are added linearly.
We can see that the secondary predictions of both D and $^3$He are in 
general agreement with the data. {\it The primary component of deuterons 
is not necessary.}

The fragmentation of heavy nuclei contribute a considerable fraction to the
deuteron flux, as already shown in \citet{2012A&A...539A..88C}. The nuclei 
heavier than helium would contribute $\sim40\%$ to D fluxes above several GeV/n,
and the nuclei between fluorine and nickel may contribute a fraction of $\sim20\%$ 
\citep{2012A&A...539A..88C}. However, for $^3$He, nuclei up to helium can 
contribute $\sim80\%$ of its flux, and nuclei up to oxygen can contribute to 
more than $90\%$ \citep{2012A&A...539A..88C}. See also Fig.~\ref{fig:dhe3_O16} 
in Appendix B the expected fluxes of D and $^3$He when setting the heaviest 
nuclei to be oxygen in the computation. It is shown that the D flux undershoots
the data by several tens percents above a few GeV/n, while the $^3$He flux is
almost consistent with the data. The main reason leading to the difference of 
heavy nuclei contribution to D and $^3$He fluxes are that the production cross 
sections for D are higher by a factor of several than those for $^3$He and T 
\citep{2012A&A...539A..88C}. For CNO nuclei, the cross section measurements 
indeed show that the D production cross sections are higher than the other two
\citep{1969ApJ...155..587R}. However, for nuclei heavier than oxygen, the 
measurement of D production cross section is lack, and the parameterization 
relies on extrapolation of light nuclei \citep{2012A&A...539A..88C}.

\begin{figure}[!htb]
\centering
\includegraphics[width=0.48\textwidth]{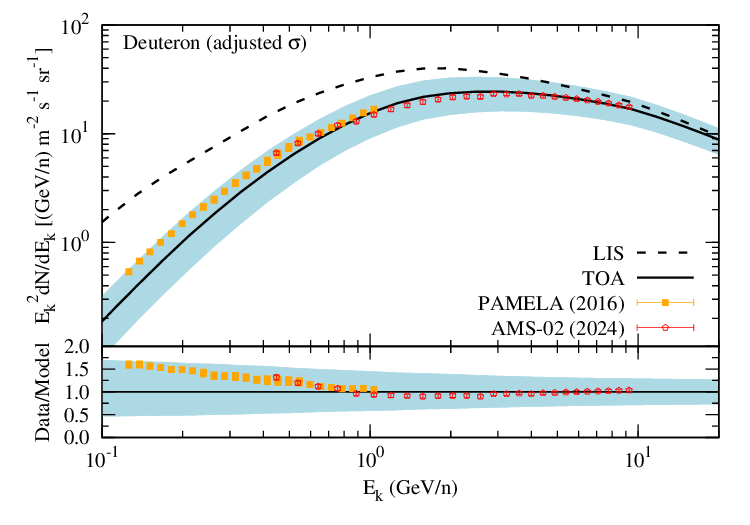}
\caption{Deuteron fluxes when adjusting slightly (see Fig.~\ref{fig:xs} 
in Appendix C) the production cross section of deuterons from 
\citet{2012A&A...539A..88C}.}
\label{fig:dxs_DRBR}
\end{figure}

From Fig.~\ref{fig:dhe3_DRBR} we also note that the predicted spectrum of 
deuterons is slightly softer than the measurement. We expect that it may 
be due to the uncertainties of the deuteron production cross sections, 
especially from nuclei heavier than C whose measurements are rare (or lack). 
As an illustration, the deuteron production cross section from the 
parameterization of \citet{2012A&A...539A..88C} for $p+{\rm C}$ interaction 
can be found in Fig.~\ref{fig:xs} of Appendix C. It is shown that the
energy coverage of the measurements is very limited, and the data points
are sparse. We thus slightly adjust the production cross section (for all 
nuclei with $Z\ge 6$) for kinetic energy above $\sim1$ GeV/n, within the
direct measurement uncertainties, in order to improve the fitting to the 
AMS-02 data. Specifically, the cross section between 1 GeV/n and 5 GeV/n
is reduced slightly, and that above 5 GeV is slightly enhanced, as shown 
by the solid line in Fig.~\ref{fig:xs}. The resulting deuteron flux is 
given in Fig.~\ref{fig:dxs_DRBR}, which matches the data well in the 
measured energy range. It is interesting to note that the adjusted cross 
section shows a better agreement with the decline trend of the measurements 
above 1 GeV/n \citep{1983PhRvC..28.1602O}. The rise above 5 GeV/n is
empirically required by the AMS-02 data, which is hard to be justified
right now since there is no measurement above 2 GeV/n. Future measurements 
of the cross sections in a wide energy range (from e.g., $O(10)$ MeV/n to 
$O(10)$ GeV/n) should be very helpful in testing such an adjustment and
the secondary origin of CR deuterons.

\begin{figure*}[!htb]
\centering
\includegraphics[width=0.48\textwidth]{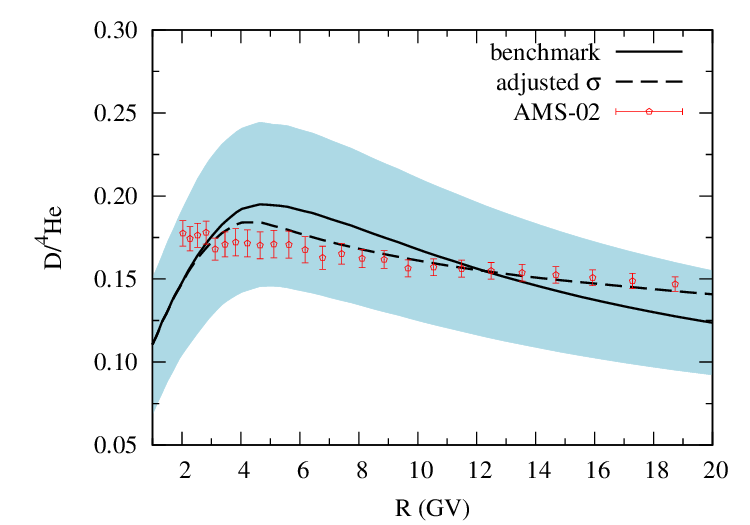}
\includegraphics[width=0.48\textwidth]{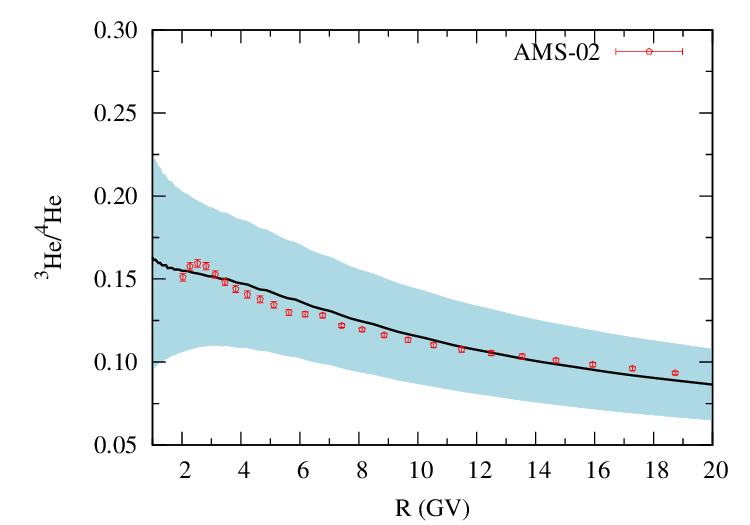}
\caption{Model predicted flux ratios of D/$^4$He (left) and $^3$He/$^4$He (right), 
compared with the AMS-02 measurements \citep{PhysRevLett.132.261001}. In the left
panel, we also show the result when adjusting slightly the production cross section
of D for $Z\ge6$ nuclei (dashed line).}
\label{fig:ratio}
\end{figure*}

We also investigate the flux ratios of D/$^4$He and $^3$He/$^4$He, 
under the framework of pure secondary origin. The results are shown in 
Fig.~\ref{fig:ratio}. Note that the ratios are measured in the rigidity 
space, and our comparison is also done in the rigidity space. Within the 
uncertainties, we can see that the model predictions are consistent with 
the data. The observed differences between the spectra of D and 
$^3$He, as well as those between the D/$^4$He (D/p) and $^3$He/$^4$He 
($^3$He/p) flux ratios, measured in the rigidity space, are expected 
to partly come from the kinetic-energy-to-rigidity conversion due to 
the different charge-to-mass ratios of D and $^3$He, in spite that 
their kinetic energy spectra do not differ much at production (some 
differences of the production spectra are expected from their different 
energy-dependent cross sections). Additional sources resulting in 
differences between D and $^3$He spectra include the propagation processes 
with different energy losses and fragmentation rates, and the solar 
modulation effect for particles with different charge-to-mass ratios.

\section{Limit on the primary contribution of deuteron}

\begin{figure}[!htb]
\centering
\includegraphics[width=0.48\textwidth]{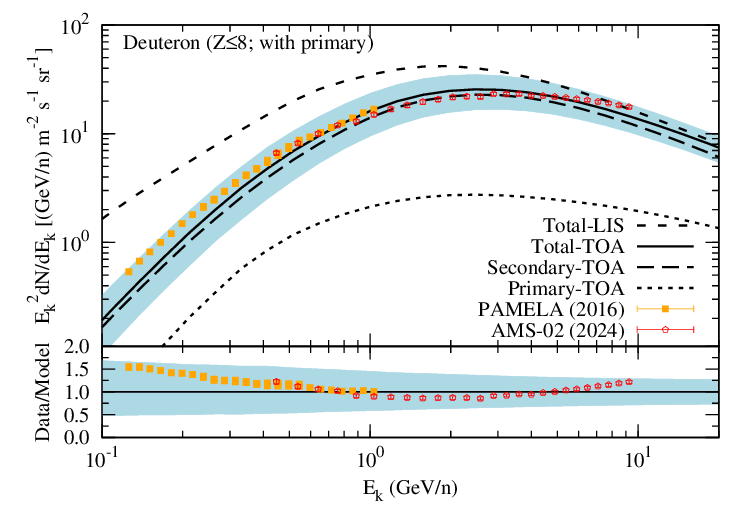}
\caption{Deuteron fluxes when adding a primary component (dotted line) with 
D/H~$=1.6\times10^{-3}$ to the secondary contribution (long dashed line) 
calculated for nuclei with $Z\leq 8$.}
\label{fig:dwp_DRBR}
\end{figure}

From the previous section, we see that a secondary origin of deuteron is 
likely to be enough to explain the measurements. Considering the possible 
uncertainties of the extrapolated deuteron production cross section, and 
also that the measured deuteron spectrum seems to be slightly harder than 
the prediction, we discuss the possibility of existence of a primary 
component of deuterons in this section. For this purpose, the heaviest 
nuclei entering the calculation is set to be oxygen, and the primary 
abundance of D/H is adjusted to match with the data. The injection spectrum 
of primary deuterons is assumed to be the same with helium. We find that 
for D/H~$=1.6\times10^{-3}$, i.e., 50 times higher than the primordial 
abundance from the Big Bang nucleosynthesis (alternatively, the acceleration 
efficiency of deuterons is about 60 times more efficient than that of protons, 
which is however hard to understand because for similar charge-to-mass ratio 
particles such as $^4$He no such strong acceleration enhancement is found), 
the model prediction is consistent with the data, as shown in 
Fig.~\ref{fig:dwp_DRBR}. In view of the results presented in 
Fig.~\ref{fig:dhe3_DRBR}, we would like to take the resulting D/H of 
$1.6\times10^{-3}$ as an upper limit on the potential primary contribution 
of deuteron. Future measurements of the deuteron spectrum in a wider energy 
range as well as the improved measurements of fragmentation cross sections 
will be crucial to testing whether there is a primary CR component of 
deuterons or not.

\section{Conclusion}

The precise measurements of spectra of secondary nuclei in CRs are very 
important in studying the propagation of CRs as well as probing new 
astrophysics or even new physics. In this work we confront the secondary 
production scenario of CR deuterons with the newest measurements by AMS-02 
\citep{PhysRevLett.132.261001}. We update the production cross sections of 
D, T, and $^3$He in the GALPROP propagation package, particularly for heavy
nuclei. We find that a CR propagation model well constrained by B/C and B/O 
ratios can consistently reproduce the data of both D and $^3$He, without 
the need to introduce primary sources of these particles. We illustrate 
that it is important to properly include the contribution of heavy nuclei 
when calculating these secondary particle yields, especially for deuterons.
The possible primary contribution of deuterons is constrained to have an 
abundance ratio of D/H to be $\lesssim 1.6\times10^{-3}$.
Further understanding of the origin of CR deuterons needs joint efforts 
in improving the measurements of fragmentation cross sections of a variety of 
channels in a wide energy range and the measurements of D and $^3$He fluxes 
towards higher energies.

\section*{Acknowledgments}
We thank L. Wu for helpful discussion. 
The ACE-CRIS data are extracted from the ACE Science Center at
http://www.srl.caltech.edu/ACE/ASC/level2/lvl2DATA\_CRIS.html.
This work is supported by the National Key Research and Development Program 
of China (No. 2022YFF0503302), the National Natural Science Foundation of China 
(No. 12220101003), the Project for Young Scientists in Basic Research of the 
Chinese Academy of Sciences (No. YSBR-061), and the New Cornerstone Science 
Foundation through the XPLORER PRIZE.

\bibliographystyle{apj}
\bibliography{refs}

\clearpage
\appendix

\setcounter{figure}{0}
\renewcommand\thefigure{A\arabic{figure}}
\setcounter{table}{0}
\renewcommand\thetable{A\arabic{table}}

\section{Propagation parameters and injection spectra of cosmic rays}

The diffusion coefficient is parameterized to be a broken power-law of 
rigidity as
\begin{equation}
D(R)=\left\{
\begin{array}{llll}
\beta^{\eta}D_0(R/R_{\rm br})^{\delta}(R_{\rm br}/R_0)^{\delta}, & {\rm for\ }R\leq R_{\rm br},\\
\beta^{\eta}D_0(R/R_{\rm br})^{\delta_h}(R_{\rm br}/R_0)^{\delta}, & {\rm for\ }R> R_{\rm br},
\end{array}
\right.
\end{equation}
where $\beta$ is the velocity of CR particles in unit of speed of light, 
$\eta$ is an empirical modification of the velocity-dependence which can 
better fit the low-energy data \citep{2010APh....34..274D}, $R_{\rm br}$ 
is the break rigidity above (below) which the slope is $\delta_h$ ($\delta$), 
and $D_0$ is the normalization at $R_0\equiv4$ GV.

The data used include:
\begin{itemize}

\item B/C: Voyager-1 \citep{2016ApJ...831...18C}, ACE, 
AMS-02 \citep{2021PhR...894....1A}, DAMPE \citep{2022SciBu..67.2162D}.

\item B/O: ACE, AMS-02 \citep{2021PhR...894....1A}, 
DAMPE \citep{2022SciBu..67.2162D}.

\item $^{10}$Be/$^9$Be: 
Ulysses \citep{1998ApJ...501L..59C}, ACE \citep{2001ApJ...563..768Y},
Voyager \citep{1999ICRC....3...41L}, IMP \citep{1988SSRv...46..205S},
ISEE-3 \citep{1988SSRv...46..205S}, and ISOMAX \citep{2004ApJ...611..892H}.

\item C \& O: 
Voyager-1 \citep{2016ApJ...831...18C}, ACE, AMS-02 \citep{2021PhR...894....1A}.

\item p \& He: 
Voyager-1 \citep{2016ApJ...831...18C}, AMS-02 \citep{2021PhR...894....1A}, 
DAMPE \citep{2019SciA....5.3793A,2021PhRvL.126t1102A}.

\end{itemize}
The Voyager-1 data are assumed to be taken out of the solar system 
\citep{2016ApJ...831...18C}, which represent the CR fluxes in the local 
interstellar environment. The other data were taken near the Earth, and 
the solar modulation needs to be included. 

\begin{figure*}[!htb]
\centering
\includegraphics[width=0.8\textwidth]{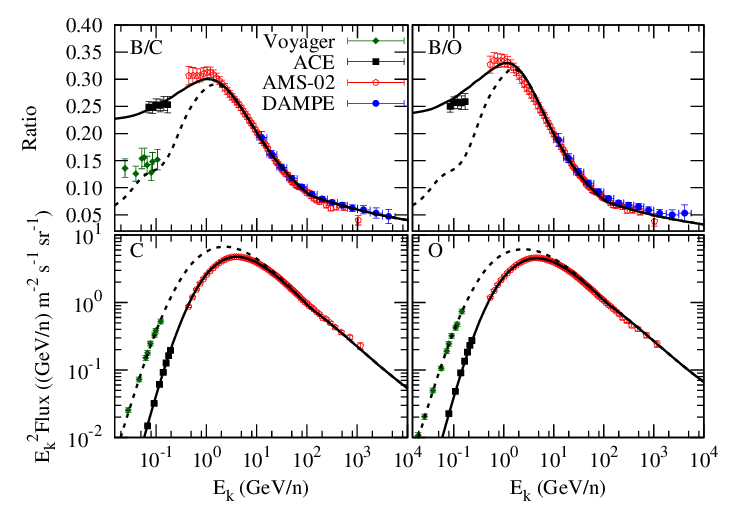}
\caption{Best-fitting results for B/C, B/O ratios and C, O fluxes. In each 
panel, the dashed line is for the LIS spectrum and the solid one is for the 
TOA spectrum with $\phi=600$ MV.}
\label{fig:BCO_DRBR}
\end{figure*}

\begin{figure*}[!htb]
\centering
\includegraphics[width=0.48\textwidth]{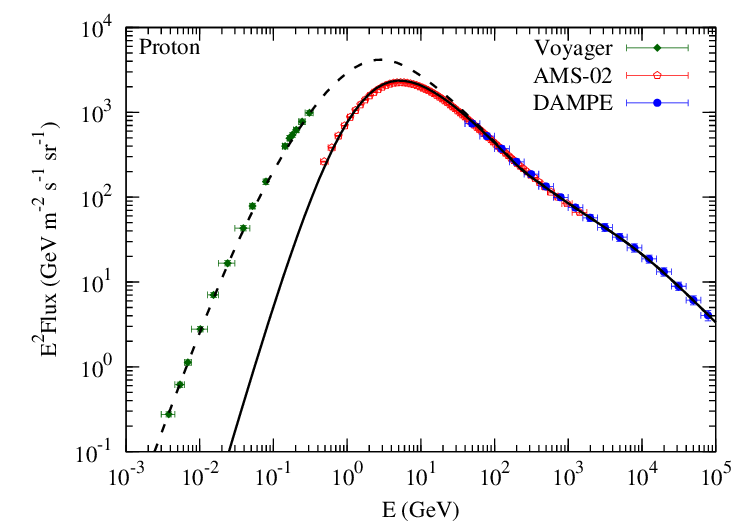}
\includegraphics[width=0.48\textwidth]{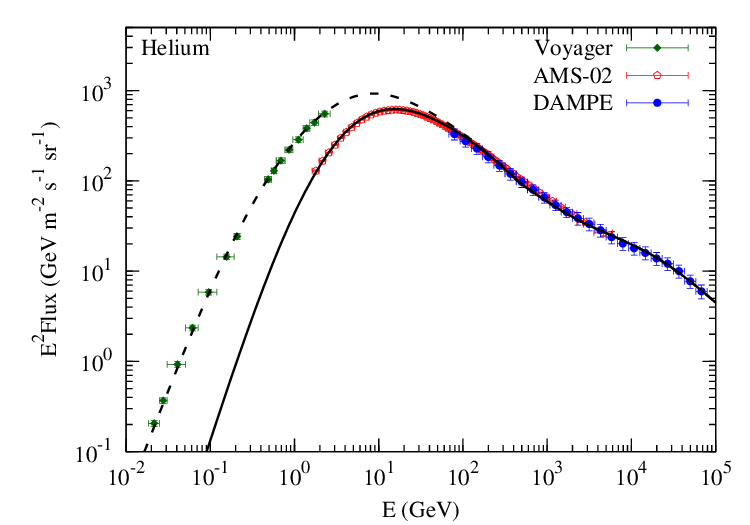}
\caption{Best-fitting results for proton and helium fluxes. In each panel, 
the dashed line is for the LIS spectrum and the solid one is for the TOA 
spectrum with $\phi=600$ MV.}
\label{fig:phe_DRBR}
\end{figure*}

The fitting propagation parameters are listed in Table \ref{tab:prop}. 
The solar modulation potential is found to be about 600 MV for the ACE and 
AMS-02 data taking time (2011-2018). The comparison between the best-fitting 
model results and the data of the above spectra and ratios are shown in 
Figs. \ref{fig:BCO_DRBR} and \ref{fig:phe_DRBR}.

\begin{table}[!htb]
\centering
\footnotesize
\caption{Best-fit values of propagation parameters.}
\begin{tabular}{cccccccc}
    \hline\hline
    $D_0$ & $\delta$ & $z_h$ & $v_A$ & $\eta$ & $R_{\rm br}$ & $\delta_h$ \\
    ($10^{28}$~cm$^2$~s$^{-1}$) & & (kpc) & (km~s$^{-1}$) & & (GV) & \\
    \hline
    $3.58$ & $0.455$ & $3.90$ & $23.5$ & $-0.72$ & $240.5$ & $0.197$ \\
    \hline
\end{tabular}
\label{tab:prop}
\end{table}

We use the measurement uncertainties of B/C and B/O ratios to characterize 
the uncertainties on the calculation of secondary nuclei fluxes due to the 
propagation model. The ACE data gives an uncertainty of $\sim5\%$ around 0.1 GeV/n. 
The uncertainties of AMS-02 data are about $5\%$ below 1 GeV/n, about $2\%\sim3\%$
between 1 and 50 GeV/n, and increase gradually to $\sim10\%$ at higher energies.
The uncertainties of DAMPE are about $5\%\sim10\%$ below 1 TeV/n, and are larger 
above 1 TeV/n. For the energy range most relevant to this work, from 0.1 GeV/n to
10 GeV/n, the uncertainties of B/C and B/O are taken to be $5\%$. The cross sections
of boron production from fragmentations of carbon, oxygen, and other nuclei will
also propagate to the uncertainties of propagation parameters. There are many
channels to produce boron, and the cross section uncertainties differ channel
by channel \citep{2018PhRvC..98c4611G}. Therefore it is difficult to quantify the 
uncertainties of boron yield due to fragmentation cross sections. According to 
several main channels, the uncertainty is estimated to be about $10\%$
\citep{2018PhRvC..98c4611G}. Combining these two effects, the uncertainty due
to propagation parameters is taken to be $15\%$. Such a value is also similar with 
those estimated in \citet{2023PhRvD.107l3008G}.

\section{Deuteron and helium-3 fluxes for fragmentation of nuclei with $Z\leq 8$}

\begin{figure*}[!htb]
\centering
\includegraphics[width=0.48\textwidth]{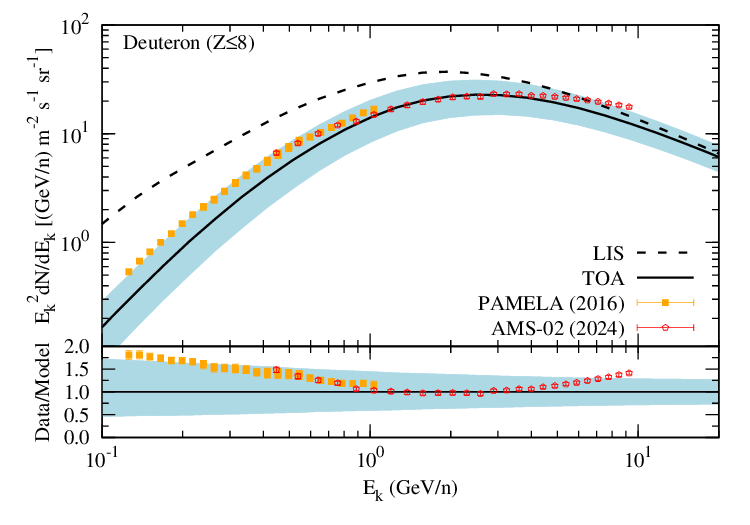}
\includegraphics[width=0.48\textwidth]{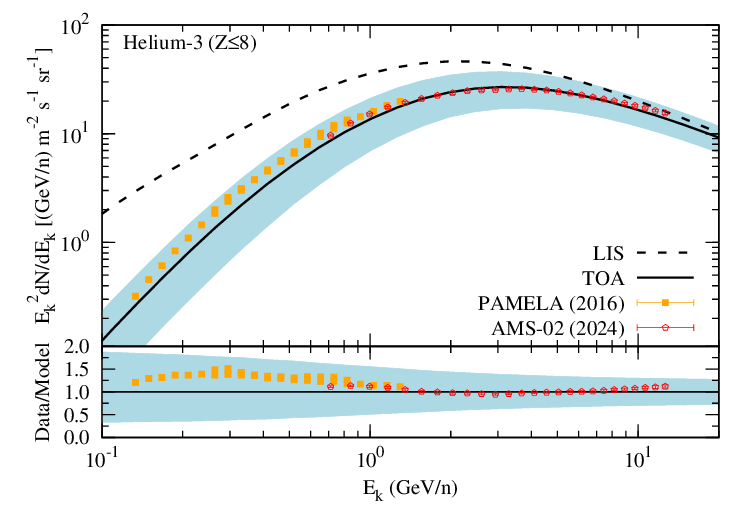}
\caption{Deuteron (left) and helium-3 (right) fluxes for fragmentation of nuclei up to 
oxygen, compared with the measurements \citep{2016ApJ...818...68A,PhysRevLett.132.261001}.}
\label{fig:dhe3_O16}
\end{figure*}

Fig.~\ref{fig:dhe3_O16} shows the results of calculated D and $^3$He fluxes 
when setting the heaviest nuclei to be oxygen in the computation. Compared with 
the data, the predicted D flux is lower at high energies, and the $^3$He flux is
consistent with the measurements. It suggests that the relative contribution of
heavy nuclei is more significant for D than for $^3$He, due to that the production 
cross sections for D are higher by a factor of several than those for $^3$He and T
\citep{2012A&A...539A..88C}.

\section{Deuteron production cross section for $p+{\rm C}$ interaction}

\begin{figure}[!htb]
\centering
\includegraphics[width=0.48\textwidth]{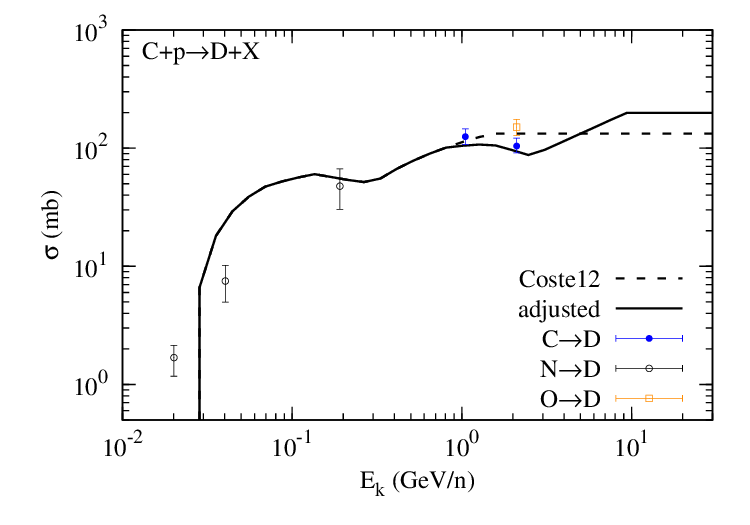}
\caption{Production cross section of deuterons from the reaction 
$p+{\rm C}\to D+X$. The dashed line is the result from the parameterization 
given in \citet{2012A&A...539A..88C}, and the solid line is the adjusted 
cross section of this work. The data of C, N, O fragmentation into D 
\citep{1969ApJ...155..587R,1983PhRvC..28.1602O}, as compiled in
\citet{2012A&A...539A..88C}, are shown. Note that, the lines should be 
compared with the blue points directly. For N and O, the normalizations 
(and energy-dependent shapes) differ slightly from those of C.}
\label{fig:xs}
\end{figure}

Fig.~\ref{fig:xs} shows the production cross section of D from the reaction 
$p+{\rm C}\to D+X$ according to the parameterization of 
\citet{2012A&A...539A..88C} (dashed line), and the adjusted one (solid line) 
used in this work. The adjustment shows a better reproduction of the decline 
trend above 1 GeV/n as shown by the measurements 
\citep[blue points;][]{1983PhRvC..28.1602O}.

\end{document}